\documentclass[]{spie}  %>>> use for US letter paper
\usepackage{amsmath,amsfonts,amssymb}
\usepackage{graphicx}
\usepackage[colorlinks=true, allcolors=blue]{hyperref}

\title{All-Fiber Broadband Wavelength-Division Multiplexing using Heterogeneous Photonic Lanterns}

\author[a]{Abani Shankar Nayak}
\author[a]{Julius Goehring}
\author[a]{Martin M. Roth}
\author[a]{Kalaga Madhav}

\affil[a]{Leibniz-Institut für Astrophysik Potsdam (AIP), An der Sternwarte 16, 14482 Potsdam,
Germany}

\authorinfo{Further author information: (Send correspondence to A.S.N.)\\A.S.N.: E-mail: anayak@aip.de, Telephone: +49 331 7499 673}

\begin{document} 
\maketitle

\begin{abstract}
This work presents an all-fiber broadband wavelength-division multiplexing architecture using $1\times7$ heterogeneous photonic lanterns (PLs), specifically $\text{PL}-{(6+1)}$ and $\text{PL}-{(4+3)}$. By combining mismatched single-mode cores, namely $\text{SM450}$ and $\text{SMF28}$, without rigid geometric constraints, these devices break system degeneracy and enable passive, wavelength-selective spatial routing. Experimental characterization reveals distinct dual-regime performance across a wide spectral range. In the visible band ($350\text{--}740\text{ nm}$), the devices function as multi-channel broadband collectors, effectively guiding light across all channels without spatial separation. Conversely, in the H-band ($1520\text{--}1610\text{ nm}$), core asymmetries lift mode degeneracy and maximize propagation constant ($\beta \, -$) matching conditions, thereby enabling passive spectral sorting. These results confirm that unconstrained, free-packaged heterogeneous fiber bundles can meet adiabatic transition criteria, showcasing their potential for next-generation astronomical facilities that require simultaneous multi-wavelength capabilities to feed separate, band-optimized spectrographs. 
\end{abstract}


\section{Introduction}
\label{sec:introduction}
The operational performance of ground-based telescopes is fundamentally limited by the Earth's atmosphere, which introduces dynamic phase distortions that degrade an incoming stellar plane wavefront into a highly aberrated profile. This distorted light cannot be efficiently coupled into a diffraction-limited single-mode fiber (SMF) without the use of complex and expensive adaptive optics (AO) systems. As a result, instruments typically employ large-core multi-mode fibers (MMFs) at the focal plane to maximize coupling efficiency, with the light then directed to spectrographs. However, MMFs come with significant drawbacks, such as severe modal noise \cite{Mahadevan_2014}, focal ratio degradation (FRD) \cite{Hernandez:21}, and scrambling instabilities \cite{Bundy_2022}, all of which compromise the wavelength calibration precision required by state-of-the-art spectrographs \cite{schmidt_2025}.

To address the architectural gap, the photonic lantern (PL) was developed as an adiabatic waveguiding device that smoothly transitions from a multi-mode waveguide input to a carefully arranged bundle of isolated single-mode cores \cite{Birks:12, Leon-Saval:10}. At the multi-mode interface, the complex, aberrated light from the telescope's focal plane is efficiently captured. As the light propagates through the longitudinal downtaper region, its chaotic spatial profile is gradually and deterministically transformed into discrete single-mode outputs \cite{Leon-Saval:10}. If the geometric transition is sufficiently gradual (adiabatic) and the number of independent single-mode ports equals or exceeds the total number of spatial modes supported at the multi-mode boundary, this modal conversion occurs essentially without loss, thereby strictly upholding the brightness theorem \cite{Birks:15}.

Traditional PL devices have been constructed as "homogeneous", utilizing an array of identical single-mode fiber cores. However, identical-fiber architectures suffer from structural degeneracies \cite{Birks:15} and do not preserve a fixed modal mapping across broad bands; consequently, a design that optimizes performance at one operational wavelength can exhibit significantly altered behavior at another \cite{Xin_2022, Taras:26}. Furthermore, a single homogeneous lantern geometry is rarely optimal across different observing regimes, as the ideal port count depends tightly on the targeted wavelength band, local wavefront quality, detector noise constraints, and the trade-off between absolute throughput and high spatial stability \cite{Jovanovic_2023}. These operational constraints make homogeneous PL highly inefficient when an instrument must dynamically switch between distinct spectral bands, varying AO performance levels, or between photon-noise- and read-noise-limited regimes \cite{Lin:21}.

To break the modal degeneracy in homogeneous lanterns, recent research has shifted toward "heterogeneous" photonic lanterns \cite{Leon-Saval:14}. The most significant recent advances focus on practical experimental demonstrations of low-loss, broadband, or mode-selective operations \cite{Francesco_2025, Becerra-Deana:25} rather than relying solely on numerical simulations \cite{Diab:21}. By utilizing mismatched cores, heterogeneous layouts lift the system's propagation constant ($\beta \, -$) degeneracy, forcing specific low-order supermodes of the multi-mode section to map preferentially onto specific single-mode cores. 

Despite recent progress, two critical research challenges remain in the development of heterogeneous photonic lanterns: \textbf{1) Broadband performance:} In the near-infrared (NIR) telecom window, a lantern's modal transmission layout changes rapidly with wavelength rather than staying completely flat. This occurs because different spatial modes travel at different phase velocities along the taper, leading to mismatched phase accumulation over the device length \cite{Dobias:26}. This problem becomes even more severe when moving down to short visible wavelengths. Because shorter wavelengths have smaller spatial features, maintaining a smooth, low-loss broadband transmission from 450 nm to 1600 nm is challenging, often resulting in high insertion losses and non-uniform transmission \cite{Zamora_2016}. \textbf{2) Fabrication tolerances:} While simulations show that mixing mismatched fiber cores works perfectly under ideal conditions, studies have also quantified how real-world manufacturing defects alter performance \cite{dana_2024}. During fabrication, errors -- such as shifts in the fiber's structural alignment or stochastic structural deformations during the capillary collapse phase -- can severely disrupt the device. Paradoxically, the very physical asymmetry used to sort the modes also makes the device highly sensitive to these geometric variations, leading to unexpected optical losses and cross-talk \cite{Wang:25}.

\begin{figure}
\begin{center}
\begin{tabular}{c} %% tabular useful for creating an array of images 
\includegraphics[width = 0.9\textwidth]{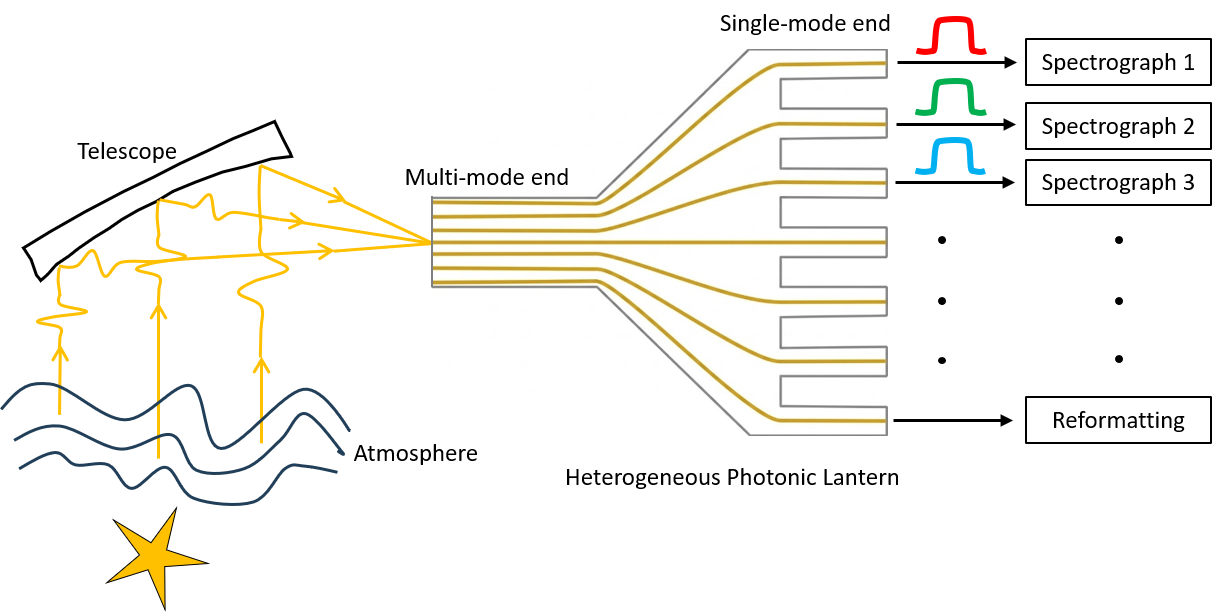}
\end{tabular}
\end{center}
\caption[] 
%>>>> use \label inside caption to get Fig. number with \ref{}
{\label{fig:concept} Conceptual schematic of a proposed astrophotonic interface that employs a heterogeneous photonic lantern to enable simultaneous multi-band astronomical observations. Starlight, disturbed by atmospheric turbulence, is focused by the telescope into a highly aberrated, multi-mode field at the focal plane. The multi-mode end of the photonic lantern captures this light and channels it into an array of isolated, diffraction-limited single-mode outputs. These stable output ends are routed to individual, band-optimized spectrographs \cite{LeonSaval_2013, Rahman_2026} or specialized reformatting channels \cite{Harris_2018, Harris_2020} for further manipulation of the starlight.}
\end{figure}

Next-generation astronomical facilities such as ANDES \cite{Marconi_2022} demand simultaneous multi-band capabilities -- capturing high-resolution visible and NIR stellar spectra to feed separate, band-optimized spectrographs. As conceptualized in Fig.~\ref{fig:concept}, an astrophotonic interface employing a heterogeneous photonic lantern can act as an all-fiber broadband wavelength-division multiplexer, eliminating the need for bulky dichroic beam splitters. Starlight, disturbed by atmospheric turbulence, is focused by the telescope into a highly aberrated, multi-mode field at the focal plane. The multi-mode end of the heterogeneous photonic lantern captures this light, breaks the modal degeneracy, and channels it into an array of isolated, diffraction-limited single-mode outputs. These stable output ends direct decoupled spectral channels into individual compact spectrographs \cite{LeonSaval_2013, Rahman_2026} or specialized reformatting systems \cite{Harris_2018, Harris_2020} for further manipulation, such as wavefront sensing, OH-suppression filtering, or aperture-masking interferometry.

To address the challenges of fabrication complexity, broadband spectral scaling, and the spatial constraints imposed by bulky dichroic beamsplitters, this proceeding aims to design, fabricate, and characterize novel heterogeneous photonic lanterns using a $1\times7$ configuration (specifically, $\text{PL}-{(6+1)}$ and $\text{PL}-{(4+3)}$) for all-fiber broadband wavelength-division multiplexing. Rather than forcing a rigid geometric configuration, this work investigates whether a randomized, free-packaged fiber bundle can reliably maintain adiabatic transition criteria and provide stable broadband spectral response. By systematically evaluating their transmission performance from visible ($350\text{--}740\text{ nm}$) to H-band ($1520\text{--}1610\text{ nm}$) spectrum, we demonstrate their potential for multi-band observations. Our results indicate that in the visible band, light is distributed across all channels simultaneously. In contrast, within the H-band, the heterogeneous core architecture enables passive spectral sorting and demultiplexing.

\section{Fabrication}
\label{sec:fabrication}

Two $1\times 7$ heterogeneous PL configurations were developed, each featuring a 1-input channel and 7-output channels. These configurations were created using a mix of single-mode fibers packaged inside low-index, custom-made, fluorine-doped silica glass capillaries produced by the Leibniz Institute of Photonic Technology (IPHT) in Jena, Germany. The capillaries have a numerical aperture (NA) of $0.11$, an inner diameter of $410 \ \mu \text{m}$, and an outer diameter of $670 \ \mu \text{m}$. To achieve multi-band operation spanning both the visible spectrum and the H-band, the devices use two distinct commercial fiber designs that were procured from Thorlabs: the \text{SM450} fiber, optimized for $488 {-} 633 \text{ nm}$, and the \text{SMF28} fiber, optimized for $1260 {–} 1625 \text{ nm}$.

The first prototype, designated $\text{PL}-(6+1)$, consists of six \text{SM450} fibers and one \text{SMF28} fiber that are randomly packed into a 7-fiber bundle. The second prototype, designated $\text{PL}-(4+3)$, features a more balanced heterogeneous distribution with four \text{SM450} fibers and three \text{SMF28} fibers. Notably, rather than using a forced, high-precision structural grid inside the capillary, both prototypes employ an unconstrained packing architecture.

Each individual fiber was cleaned with isopropanol alcohol (IPA) to remove dust and surface residues. The input fibers were carefully stripped in designated sections and then inserted into the custom-made glass capillary, which was also cleaned with IPA. A vacuum pump was connected to the capillary to eliminate any remaining IPA, as residual solvent could interfere with the subsequent tapering process. After the preparation, the loaded capillary was mounted in the \textit{Vytran GPX-3000} tapering system. During tapering, the capillary and the enclosed fibers were locally softened and drawn into a unified structure known as the PL, where the individual fibers formed the guiding cores and the capillary served as the surrounding cladding. The taper consists of three regions: a downtaper, where the inner diameter decreases from $410 \ \mu \text{m}$ to $50 \ \mu \text{m}$; a uniform waist section of $50 \ \mu \text{m}$; and an uptaper, where the inner diameter increases back to $410 \ \mu \text{m}$. A long, optimized downtaper length of $4 \text{ cm}$ \cite{Davenport:21, Rypalla_2024} is used across all configurations to maintain adiabatic conditions in the PL. Following tapering, the PL was transferred to the \textit{Vytran LDC-400} cleaver. The device applied controlled tension to the PL, and a diamond blade repeatedly contacted the PL at the waist section until a crack initiated and propagated cleanly through the structure. This process produced a flat, high-quality end facet suitable for splicing. To complete the device, the cleaved PL was spliced to a multi-mode output fiber with a core diameter of $50 \ \mu \text{m}$ (Thorlabs, \text{FG050LGA}) using the hand-held fusion splicer, \textit{Sumitomo Electric Type-Q101-CA+}. Additionally, the heterogenous PL is securely glued into the groove of a custom-made 3D-printed plate (see \textit{Right Inset} of Fig.~\ref{fig:setup}). This plate provides critical mechanical protection for the fragile tapered and spliced regions of the lantern. 

\section{Experimental Setup}
\label{sec:experimental_setup}

\begin{figure}
\begin{center}
\begin{tabular}{c} %% tabular useful for creating an array of images 
\includegraphics[width = 0.9\textwidth]{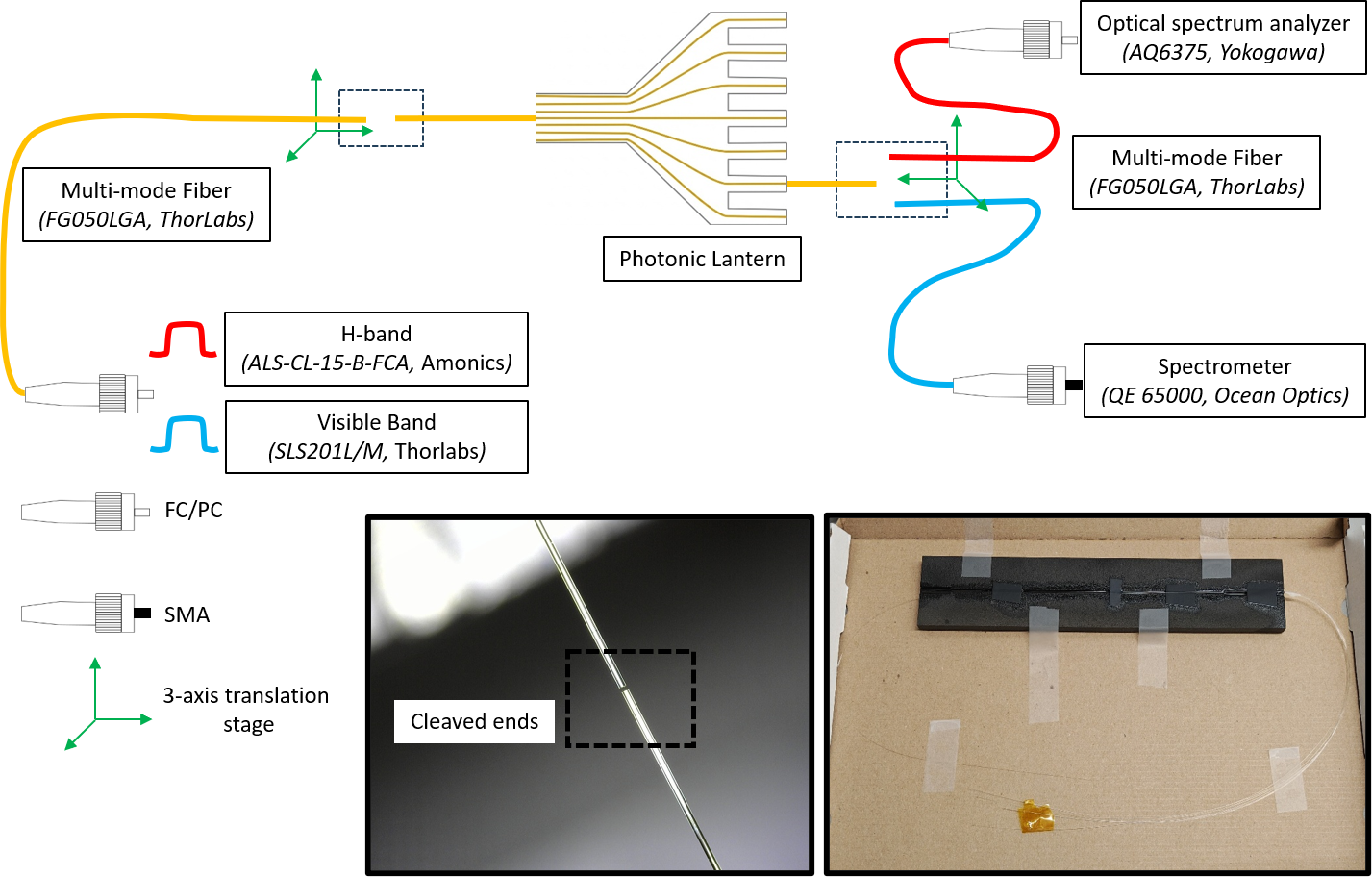}
\end{tabular}
\end{center}
\caption[] 
%>>>> use \label inside caption to get Fig. number with \ref{}
{\label{fig:setup} A schematic of the experimental setup used for evaluating multi-band performance of $1\times7$ heterogeneous photonic lanterns. This setup features dual-wavelength launch and collection capabilities. High-precision three-axis $\text{XYZ}$ translation stages are employed to optimize the butt-coupling alignment at both the input multi-mode ends and the individual single-mode ends of the PL. The collected light is selectively directed into either an H-band-optimized optical spectrum analyzer (indicated by the red path) or a visible-spectrum-optimized spectrometer (indicated by the blue path). \textit{Left Inset:} A micrograph displays the high-quality cleaved bare ends of the PL and the fiber during the alignment process. A clean, flat cleave is essential for minimizing coupling losses in the setup. \textit{Right Inset:} A photograph shows the $1\times7$ heterogeneous PL, which is securely glued into the groove of a custom-made 3D-printed plate. This plate provides critical mechanical protection for the fragile tapered and spliced regions of the lantern.}
\end{figure}

To evaluate the multi-band performance of the fabricated $\text{PL}-{(6+1)}$ and $\text{PL}-{(4+3)}$ configurations, a specialized optomechanical testbed was established as illustrated in Fig.~\ref{fig:setup}. To assess the full operational bandwidth of the devices, two separate optical excitation sources were deployed at the injection interface: \textbf{a)} For the visible band: A stable, broadband halogen lamp (\textit{SLS201/M, Thorlabs}) that provides continuous spectral illumination. \textbf{b)} For the H-band: A high-power Amplified Spontaneous Emission-based source (\textit{ALS-CL-15-B-FCA, Amonics}) that delivers uniform, high-dynamic-range illumination. The chosen illumination source is coupled into an injection multi-mode fiber with a core diameter of $50 \ \mu \text{m}$ (\textit{FG050LGA, Thorlabs}). The other end of the multi-mode fiber was cleaved and mounted on a high-precision, three-axis $\text{XYZ}$ translation stage. This stage enables reproducible, stable alignment for butt coupling to the multi-mode end of the PL. 

To systematically scan each individual single-mode channel of the PL, a secondary three-axis $\text{XYZ}$ translational is positioned at the output interface. This stage is equipped with another cleaved multi-mode fiber end with a core diameter of $50 \ \mu \text{m}$ (\textit{FG050LGA, Thorlabs}) to sequentially collect light from the individual channels of the PL. As shown in Fig.~\ref{fig:setup}, depending on the source used during the injection, the output collection fiber directs the optical power to two specialized detectors: \textbf{a)} The H-band light (Red path) is directed to an Optical Spectrum Analyzer (\textit{AQ6375, Yokogawa}). \textbf{b)} The Visible band light (Blue path) is transmitted to a highly sensitive spectrometer (\textit{QE65000, Ocean Optics}). 

During all measurement sequences, a dark frame is recorded to eliminate ambient light and instrumentation noise. To achieve calibrated transmission values, a bare multi-mode fiber (\textit{FG050LGA, Thorlabs}) with a core diameter of $50\ \mu\text{m}$ and a length of $1\ \text{m}$ is used for the reference measurement. Both ends of this fiber are cleaved so that it can serve as a substitute for the PL during the reference measurement. This reference measurement compensates for spectral ripple or non-uniformities in the illumination source outputs, enabling clean, linear normalization across all evaluated PL channels.

\section{Results of the $\text{PL}-(6+1)$ configuration}
\label{sec:results_PL_6+1}

\begin{figure}
   \begin{center}
   \begin{tabular}{cc} %% tabular useful for creating an array of images 
   \includegraphics[width = 0.45\textwidth]{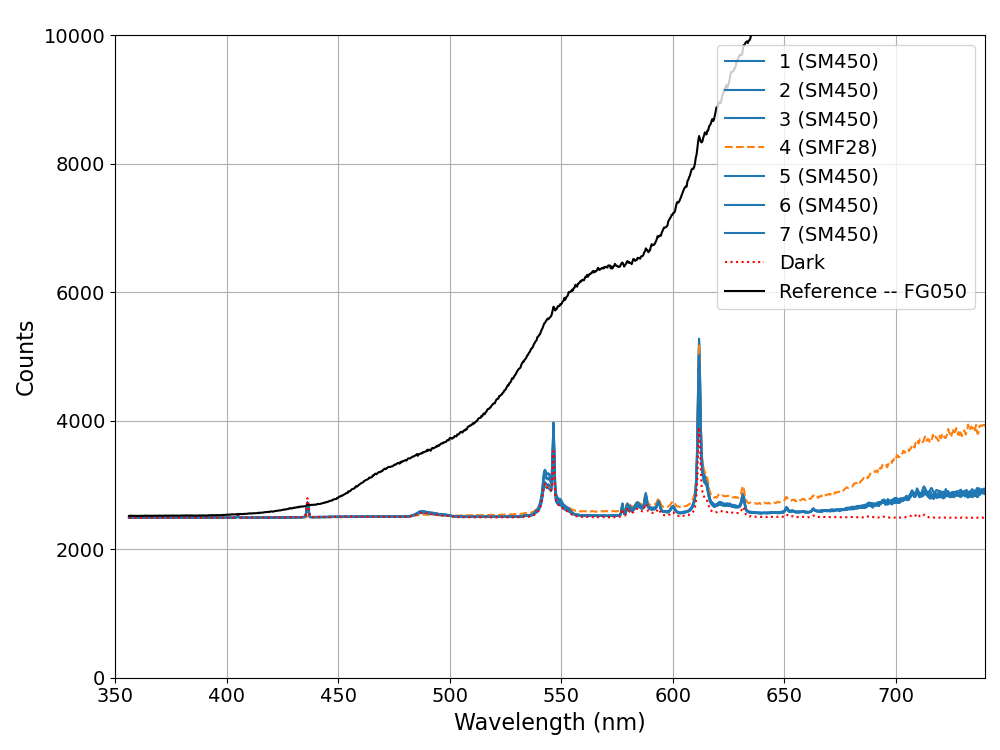} &
   \includegraphics[width = 0.45\textwidth]{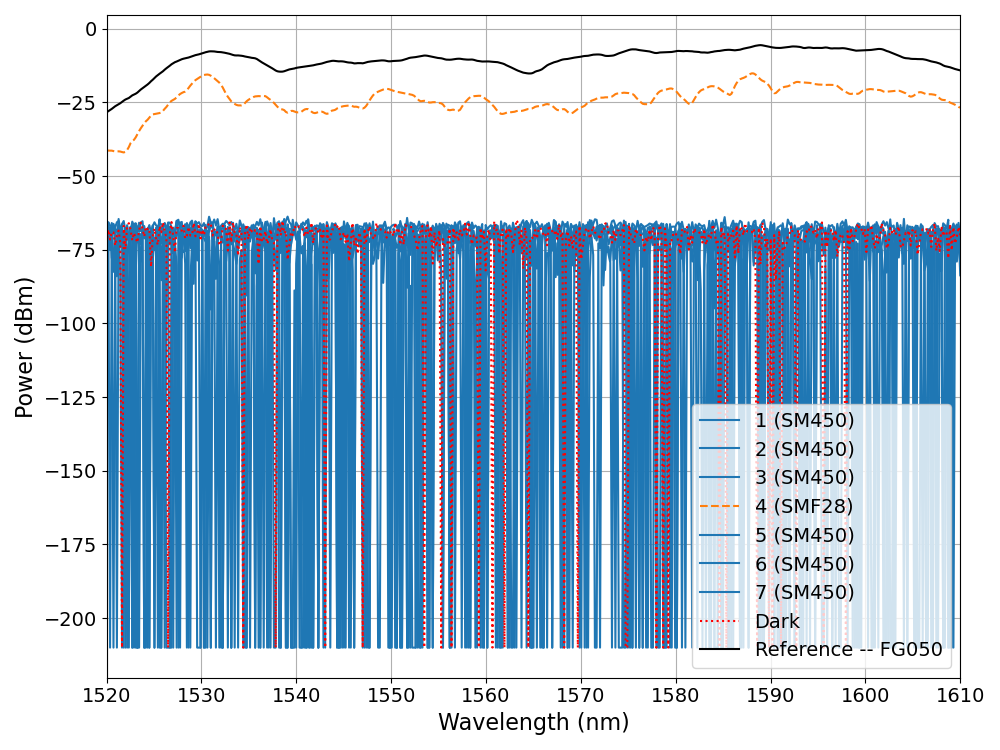}\\
   \small(a) & \small(b) 
   \end{tabular}
   \end{center}
   \caption[] 
%>>>> use \label inside caption to get Fig. number with \ref{}
   {\label{fig:results_raw_6+1} Raw spectral responses of the $\text{PL}-{(6+1)}$ configuration characterized across two distinct regimes: (a) Visible band ($350\text{--}740\text{ nm}$) captured using a spectrometer, and (b) H-band ($1520\text{--}1610\text{ nm}$) measured on an OSA. The solid black line denotes the $50\ \mu\text{m}$ multi-mode core reference fiber (Thorlabs, $\text{FG050LGA}$) of length $1\text{ m}$. Channel 4, which is randomly assigned, corresponds to the single $\text{SMF28}$ fiber located within the 7-fiber bundle matrix.}
   \end{figure}

The raw spectral profiles for the $\text{PL}-{(6+1)}$ photonic lantern configuration are presented in Fig.~\ref{fig:results_raw_6+1}. This configuration consists of six $\text{SM450}$ fibers and one $\text{SMF28}$ fiber bundled together within a capillary matrix. Notably, the $\text{SMF28}$ fiber, which was naturally integrated into the bundle's cross-section, is randomly designated as Channel 4. In the visible band (Fig.~\ref{fig:results_raw_6+1}(a)), the transmission behavior clearly reflects the underlying waveguide cutoff conditions. Below $620\text{ nm}$, the throughput across all ports is severely limited by system coupling constraints, the source profile, and an instrumentation signal that approaches the dark background level. Beyond $620\text{ nm}$, the optical throughput within the $\text{SMF28}$ fiber increases significantly as its core begins to support multi-mode propagation at these shorter wavelengths. In contrast, the signals from the six $\text{SM450}$ channels exhibit an exceptionally uniform profile, displaying nearly identical spectral shapes and intensities. However, their absolute counts remain lower than those of the $\text{SMF28}$ channel. In the H-band (Fig.~\ref{fig:results_raw_6+1}(b)), the infrared light is almost entirely transmitted through the single $\text{SMF28}$ fiber. In stark contrast, the transmission from the six $\text{SM450}$ channels drops to the instrumentation noise floor at approximately $-75\text{ dBm}$. This indicates a high-dynamic-range port isolation of about $50\text{ dB}$ relative to the active $\text{SMF28}$ channel, which operates around $-20\text{ dBm}$. The dominance of a single channel for infrared transmission confirms that even when randomly distributed within an unconstrained bundle, the larger core profile of the $\text{SMF28}$ fiber acts as the primary guiding channel for infrared light.

\begin{figure}
   \begin{center}
   \begin{tabular}{cc} %% tabular useful for creating an array of images 
   \includegraphics[width = 0.45\textwidth]{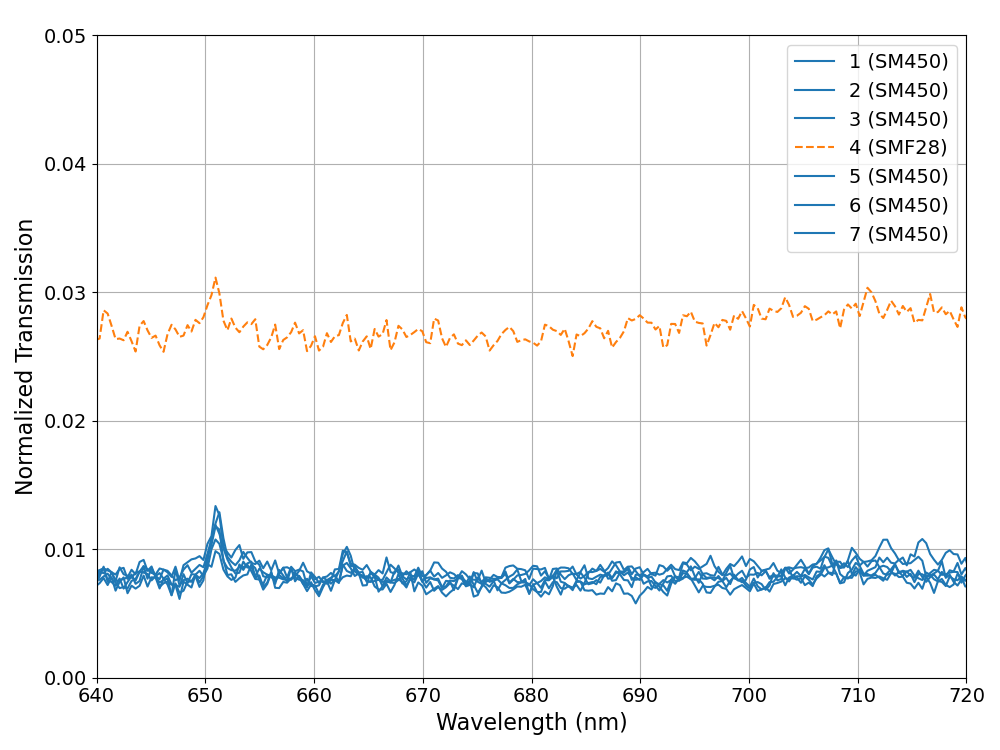} &
   \includegraphics[width = 0.45\textwidth]{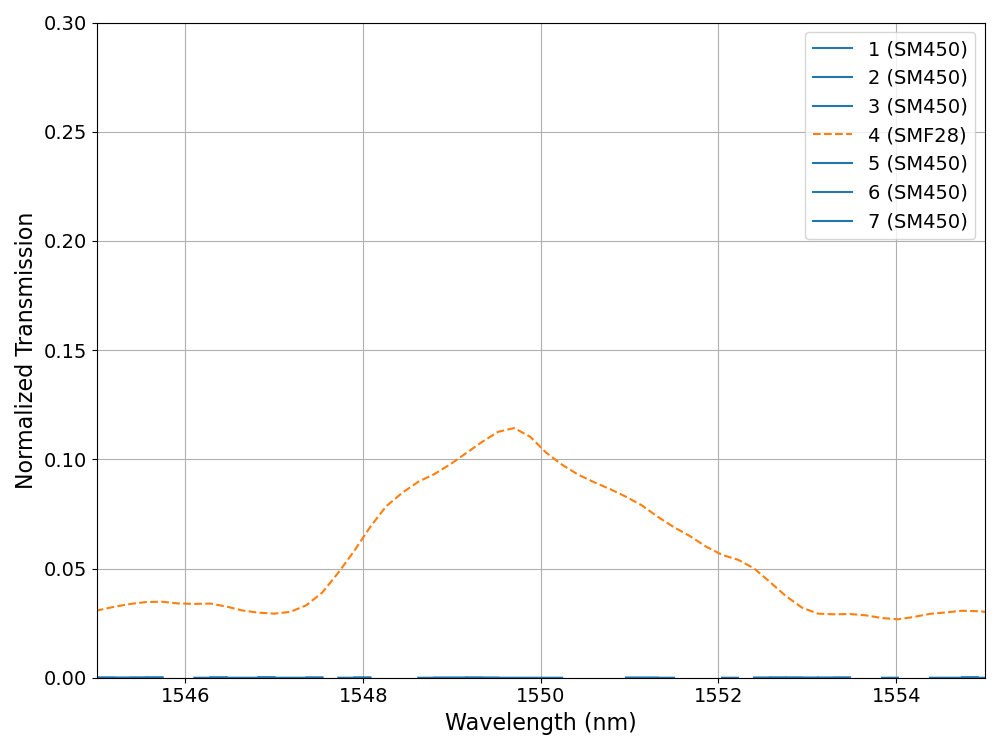}\\
   \small(a) & \small(b) 
   \end{tabular}
   \end{center}
   \caption[] 
%>>>> use \label inside caption to get Fig. number with \ref{}
   {\label{fig:results_normalized_6+1}  Dark-corrected and normalized transmission efficiencies of the $\text{PL}-{(6+1)}$ configuration calibrated against the $50\ \mu\text{m}$ core multi-mode fiber reference: (a) Visible band ($640\text{--}720\text{ nm}$), and (b) H-band ($1545\text{--}1555\text{ nm}$). To improve visualization and eliminate edge artifacts, the x-axes are cropped strictly to the regions of interest where the reference spectrum profile exhibits a flat profile (see Fig.~\ref{fig:results_raw_6+1}). Discontinuities in all $\text{SM450}$ channels in the H-band are represented as NaN values due to negative values after normalization.}
   \end{figure}

To quantify the transmission metrics, the raw datasets from Fig. \ref{fig:results_raw_6+1} were dark-corrected and linearly normalized against a reference spectrum obtained from a $50\ \mu\text{m}$ core bare multi-mode fiber (Thorlabs, $\text{FG050LGA}$) with a length of $1\text{ m}$. The reference was taken under identical launch conditions. The resulting normalized spectral profiles are shown in Fig.~\ref{fig:results_normalized_6+1}. Importantly, the spectral windows displayed in the figures are truncated to specific regions of interest (ROI): $640\text{--}720\text{ nm}$ for the visible band and $1545\text{--}1555\text{ nm}$ for the H-band. This cropping focuses on the bands where the reference light source exhibits a relatively flat profile, helping to eliminate unwanted oscillations in the signal and facilitating a better understanding of the underlying physics. In the visible window (Fig. \ref{fig:results_normalized_6+1}(a)), the device shows a remarkably flat performance profile. The single $\text{SMF28}$ channel maintains a uniform transmission efficiency with a mean value of approximately 0.027 $(2.7\%)$. Meanwhile, the remaining six $\text{SM450}$ channels cluster closely together with a mean transmission efficiency of about 0.008 $(0.8 \%)$. The standard deviations are below 0.0008, indicating exceptional uniformity across all six $\text{SM450}$ channels and highlighting a highly symmetric preservation of the cross-section during the tapering process. In the H-band window (Fig.~\ref{fig:results_normalized_6+1}(b)), the normalized data reveals the distinct spectral sorting of the device. Within this flat ROI, the $\text{SMF28}$ channel shows a peak efficiency of approximately $11.5\%$ at around $1550\text{ nm}$. The mean and median efficiencies for this channel are $5.6\%$ and $4.1\%$, respectively. This localized peak may result from the highly efficient $\beta \,-$ matching condition of the fundamental mode at $1550\text{ nm}$, due to the larger core size of the $\text{SMF28}$ fiber compared to the $\text{SM450}$ fiber. In contrast, the remaining six $\text{SM450}$ channels demonstrate an absolute transmission efficiency of roughly $0\%$ on a linear scale, confirming complete power rejection and excellent cross-talk isolation throughout the unconstrained tapered region of the PL. It is important to note that the minor gaps or missing data segments in all the $\text{SM450}$ channels are due to numerical artifacts that arise from normalizing with the reference fiber. Low-signal data points near the instrumental noise and negative values were systematically converted to NaN (Not-a-Number) values to eliminate spurious fluctuations and ensure clearer data visualization.

\section{Results of the $\text{PL}-(4+3)$ configuration}
\label{sec:results_PL_4+3}

\begin{figure}
   \begin{center}
   \begin{tabular}{cc} %% tabular useful for creating an array of images 
   \includegraphics[width = 0.45\textwidth]{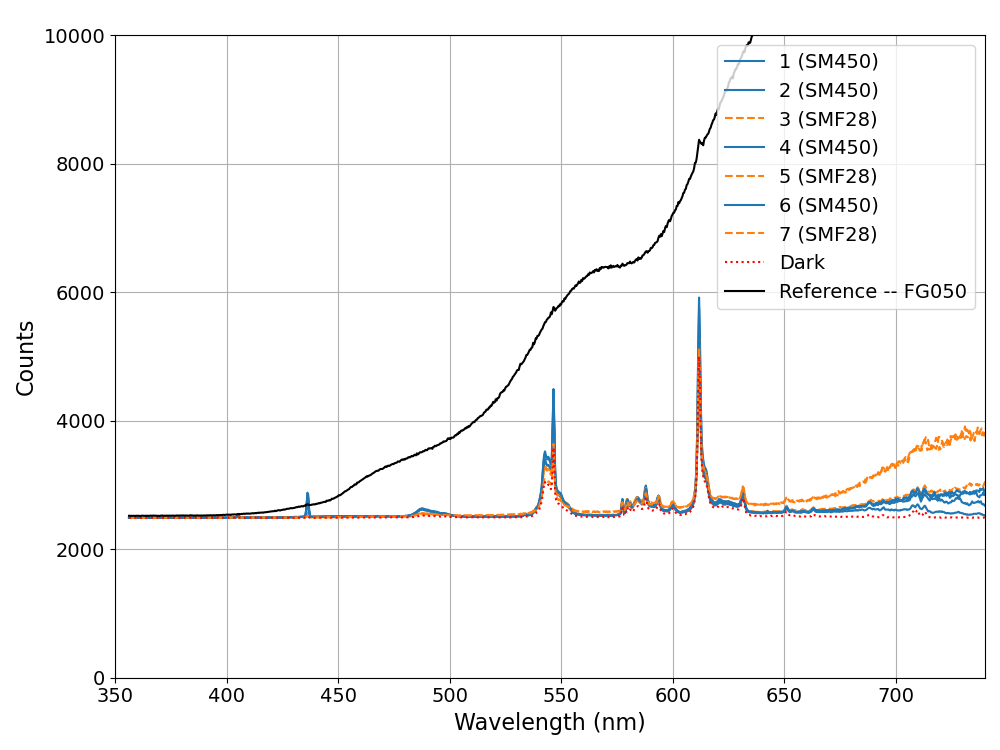} &
   \includegraphics[width = 0.45\textwidth]{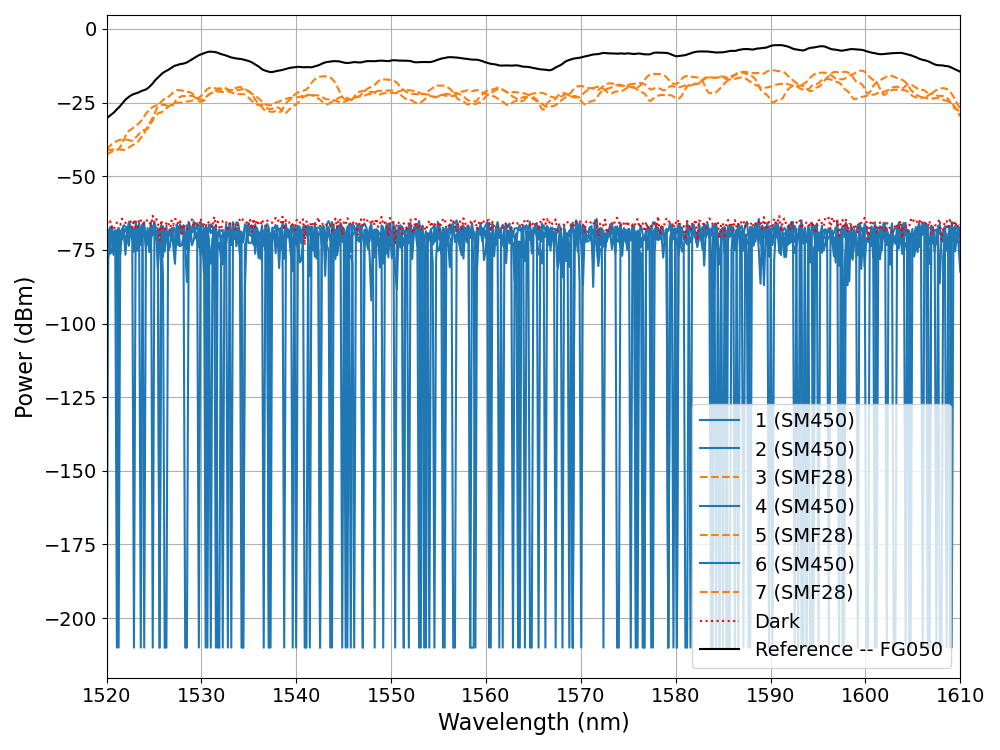}\\
   \small(a) & \small(b) 
   \end{tabular}
   \end{center}
   \caption[] 
%>>>> use \label inside caption to get Fig. number with \ref{}
   {\label{fig:results_raw_4+3} Raw spectral responses of the $\text{PL}-{(4+3)}$ configuration characterized across two distinct regimes: (a) Visible band ($350\text{--}740\text{ nm}$) captured using a spectrometer, and (b) H-band ($1520\text{--}1610\text{ nm}$) measured on an OSA. The solid black line represents the $50\ \mu\text{m}$ multi-mode core reference fiber (Thorlabs, $\text{FG050LGA}$) with a length of $1\text{ m}$. The channels are assigned randomly. Channels numbered 1, 2, 4, and 6 correspond to $\text{SM450}$ fibers, while channels 3, 5, and 7 are represented by $\text{SMF28}$ fibers within the 7-fiber bundle matrix. In the H-band, several data points for the $\text{SM450}$ channels are completely missing. This discontinuity occurs where the low signal falls below the instrumentation noise (indicated in red), resulting in negative or NaN values. These values were systematically omitted to exclude spurious fluctuations.}
   \end{figure}

The raw spectral profiles for the $\text{PL}-{(4+3)}$ configuration are presented in Fig.~\ref{fig:results_raw_4+3}. The channel assignments are randomly distributed across the cross-section, with channels 1, 2, 4, and 6 corresponding to the $\text{SM450}$ fiber and channels 3, 5, and 7 assigned to the $\text{SMF28}$ fiber. In the visible spectrum (Fig.~\ref{fig:results_raw_4+3}(a)), the transmission profiles align with the waveguide cutoff conditions, similar to those shown in Fig.~\ref{fig:results_raw_6+1}. Beyond $620$ nm, we observe a divergence in behavior among the individual cores. Two of the $\text{SMF28}$ channels -- 5 and 7 -- demonstrate the highest throughput, rapidly exceeding 1000 counts. This increased transmission is due to their larger core size, which supports multi-mode propagation at these wavelengths. Interestingly, the third $\text{SMF28}$ channel 3 follows an intermediate path, showing lower intensities than channels 5 and 7 but still distinctly above the cluster of $\text{SM450}$ channels. This spectral variance may suggest uneven local modal excitation within the fused cross-section. Additionally, three out of the remaining four $\text{SM450}$ channels form a tightly grouped cluster with readings below 500 counts. When transitioning to the H-band (Fig.~\ref{fig:results_raw_4+3}(b)), the behavior of core-dependent guidance changes dramatically. Within this range, all three SMF28 channels (3, 5, and 7) cluster closely together, maintaining high power levels between $-15\text{ dBm}$ and $-25\text{ dBm}$. This tight grouping indicates that the three $\text{SMF28}$ cores collectively serve as a primary spatial confinement zone for the infrared light. In contrast, the four \text{SM450} channels experience significant suppression, dropping below the instrumentation noise floor to $-67\text{ dBm}$. This substantial vertical separation translates to a raw channel isolation of $\sim 47 \text{ dB}$, similar to the value obtained in Sec.~\ref{sec:results_PL_6+1}. It is important to mention that several data points for the SM450 channels are completely missing; this discontinuity occurs when the low signal falls below the instrumentation noise level, resulting in negative or NaN values. These values were systematically excluded to prevent spurious fluctuations in the data visualization.

\begin{figure}
   \begin{center}
   \begin{tabular}{cc} %% tabular useful for creating an array of images 
   \includegraphics[width = 0.45\textwidth]{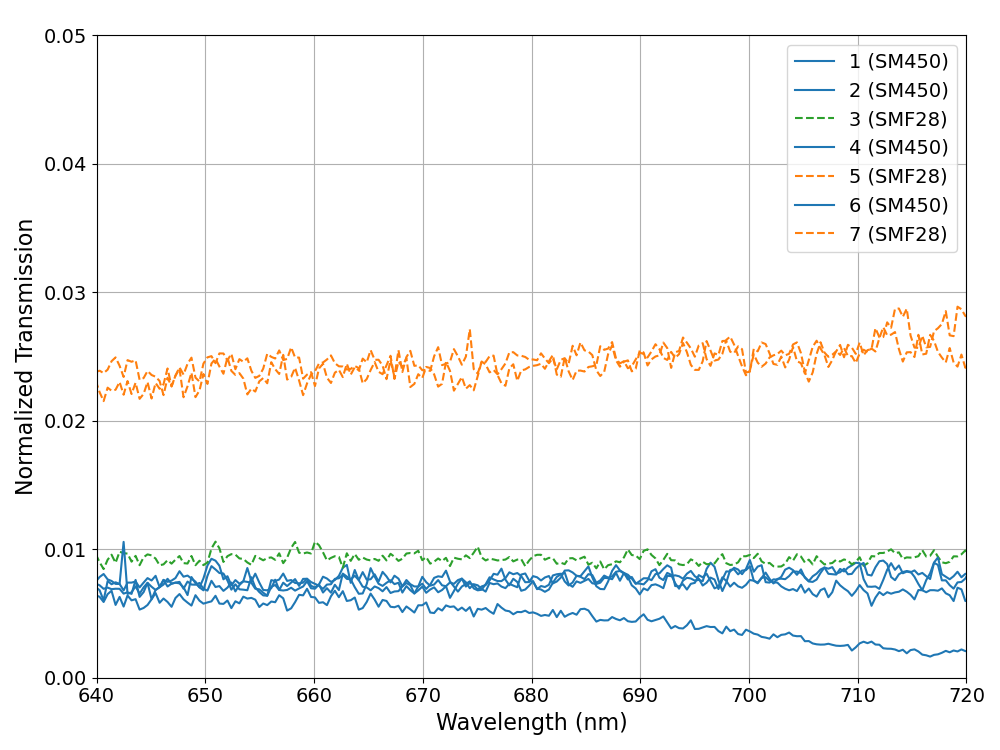} &
   \includegraphics[width = 0.45\textwidth]{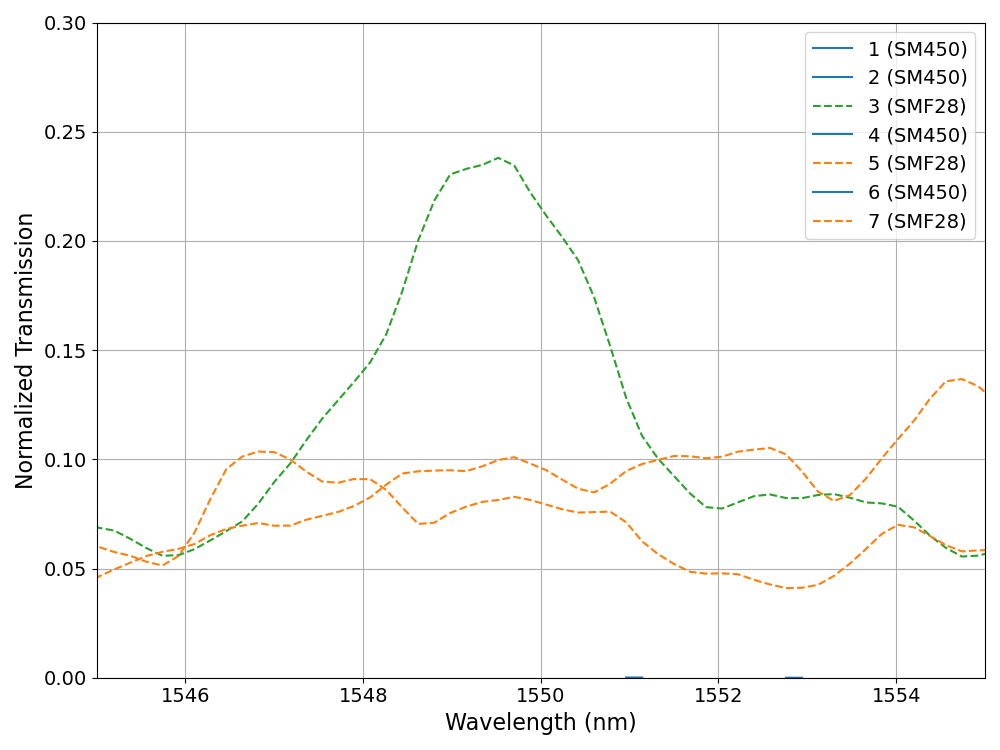}\\
   \small(a) & \small(b) 
   \end{tabular}
   \end{center}
   \caption[] 
%>>>> use \label inside caption to get Fig. number with \ref{}
   {\label{fig:results_normalized_4+3} Dark-corrected and normalized transmission efficiencies of the $\text{PL}-{(4+3)}$ configuration calibrated against a $50\ \mu\text{m}$ core multi-mode fiber reference: (a) Visible band ($640\text{--}720\text{ nm}$), and (b) H-band ($1545\text{--}1555\text{ nm}$). The x-axes are cropped strictly to the flat region of interest of the reference source. Discontinuities along the $\text{SM450}$ channels in the H-band represent NaN values due to negative values after normalization.}
   \end{figure}

The dark-corrected and normalized spectral transmission profiles for the $\text{PL}-{(4+3)}$ configuration are presented in Fig.~\ref{fig:results_normalized_4+3}. To improve clarity, the spectral windows have been cropped to focus on the regions of interest (ROI): $640\text{--}720\text{ nm}$ for the visible band and $1545\text{--}1555\text{ nm}$ for the H-band, where the reference spectrum is particularly flat. In the visible window (Fig.~\ref{fig:results_normalized_4+3}(a)), the normalized throughput profiles show a distinct distribution of power across the unconstrained fiber bundle. Channels 5 and 7, which utilize the $\text{SMF28}$ fibers, achieve mean efficiencies of $2.4\%$ and $2.5\%$, respectively, placing them in the upper transmission range. In contrast, all channels using $\text{SM450}$ fibers have mean efficiencies between $0.5\%$ and $0.8\%$. A notable finding is observed in Channel 3 ($\text{SMF28}$), which is distinct from the upper transmission bands of channels 5 and 7. Channel 3 is situated just above all the lower $\text{SM450}$ channels, maintaining a stable mean efficiency of $0.9\%$.

When evaluating the H-band window (Fig. \ref{fig:results_normalized_4+3}(b)), we observe a total inversion in behavior. As expected, all four $\text{SM450}$ channels are heavily suppressed, dropping to the instrumentation noise threshold, resulting in NaN statistical values and missing data in the figure. In contrast, optical waveguiding is maintained exclusively by the three $\text{SMF28}$ channels, which show highly localized, wavelength-dependent peaks. Remarkably, channel 3 stands out as the most efficient channel in the entire device, producing a significant transmission profile that peaks at $23.8 \%$ at a wavelength of $1549.52 \text{ nm}$. A similar peak was previously observed in Fig. \ref{fig:results_normalized_6+1} around $1550 \text{ nm}$. This consistent central peak is thought to arise from the physical dynamics of the adiabatic taper rather than numerical normalization, as the analysis window was intentionally restricted to a flat spectral profile. This localized behavior suggests that the evolving low-order supermodes \cite{Leon-Saval:14} of the multi-mode cross-section map onto the spatial arrangement of the active $\text{SMF28}$ cores, with a maximum $\beta \, -$ overlap centered at $1550 \text{ nm}$.

The counterintuitive finding -- where channel 3 is heavily suppressed in the visible range but highly dominant in the infrared -- provides insight into the internal physics of unconstrained, randomly packed photonic lanterns. During fabrication, the fiber bundles are organically packed, and channel 3 may occupy a central, high-symmetry position within the capillary cross-section. This can be verified only by measuring the refractive index cross-section of the tapered region or by replicating a similar configuration. Nevertheless, this striking contrast illustrates that a randomized, unconstrained fabrication process can still yield high-efficiency, mode-selective \cite{Leon-Saval:14} routing driven purely by waveguide geometry.

\section{Conclusion}
\label{sec:conclusion}

In this proceeding, we have demonstrated a novel architectural approach to all-fiber broadband wavelength-division multiplexing utilizing heterogeneous photonic lanterns. By mixing highly mismatched single-mode fiber cores ($\text{SM450}$ and $\text{SMF28}$) within an unconstrained capillary packaging pipeline, we successfully fabricated and characterized two $1\times7$ heterogeneous PL architectures, namely $\text{PL}-{(6+1)}$ and $\text{PL}-{(4+3)}$. 

Experimental characterization of the fabricated lanterns reveals a compelling dual-regime optical performance across a wide spectral range: \textbf{In the Visible spectrum}, the devices operate as multi-channel broadband collectors, uniformly guiding light across all output channels without spatial separation. \textbf{In the H-band spectrum}, the core asymmetries successfully break system degeneracy, maximizing the $\beta \,- $ matching conditions to enable passive spectral sorting, thereby allowing for wavelength-division multiplexing exclusively in the infrared rather than the visible regime.

Notably, the $\text{PL}-{(4+3)}$ configuration demonstrates unique spectral routing dynamics that highlight the advantages of balancing the spatial distribution of mismatched fiber pairs. An intriguing finding is that one of the channels is significantly suppressed in the visible range yet highly dominant in the infrared. This insight suggests that a randomized, unconstrained fabrication process can still achieve high-efficiency, mode-selective routing, driven solely by the waveguide geometry.

By simultaneously enabling broadband visible waveguiding and passive infrared spectral demultiplexing within a single integrated all-fiber component, these heterogeneous photonic lanterns could be incorporated into next-generation astronomical facilities \cite{Marconi_2022}. Future work will focus on repeatability and stability, and on quantifying the precise manufacturing tolerances of these free-packaged bundles to further optimize the multi-band spectral profiles across the visible and infrared regions.

\section{Acknowledgments}
The authors would like to express their sincere gratitude to Dr. Adrian Lorenz, Dr. Tobias Habisreuther, and Dr. Jörg Bierlich, all from IPHT, for fabricating the custom capillary tubes used in this study. We also extend our thanks to Mr. Dennis Plüschke (AIP) for his meticulous execution of the fiber-gluing process. Additionally, we are grateful to Mr. Svend-Marian Bauer (AIP) for designing the 3D-printed protective plate, which provided essential mechanical support for the photonic lanterns. We gratefully acknowledge financial support for this work from the PICS4SENS project, funded by the State of Brandenburg through the Investitionsbank des Landes Brandenburg (ILB), with support from the European Regional Development Fund (ERDF/EFRE), grant number 86000879.

\section{Declaration of AI use}
During the preparation of this work, the authors used Grammarly and Gemini 3.5 Flash to improve the language, readability, and grammatical structure of the manuscript. After using these tools, the authors reviewed, verified, and edited the content as needed, and take full responsibility for the final text of the publication.

% References
\bibliography{report} % bibliography data in report.bib
\bibliographystyle{spiebib} % makes bibtex use spiebib.bst

\end{document}